# Arthur Neville Brown: schoolmaster and variable star observer

**Jeremy Shears**

**Abstract**

Arthur Neville Brown, MA, FRAS (1864-1934) was a prolific variable star observer and served for many years as Secretary of the BAA Variable Star Section. This paper discusses Brown's life and career, both as a variable star enthusiast and as a dedicated and highly respected schoolmaster.

**Introduction**

The reputation of the BAA as one of the world's leading organisations for observational astronomy is largely due to the dedication of its members, who go out night after night to secure observations, and its Section Directors, Officers and Council who provide the necessary strategic direction and infrastructure. In any large organisation, effective and visible leadership is essential for it to meet its objectives and the BAA is no different. Over the years the Association has benefited from the leadership provided by stalwarts such as T.H.E.C. Espin, T.E.R. Phillips, and W.H. Steavenson, whose names are celebrated even today. However, the contributions of many other less well-known people, who often work quietly behind the scenes, are equally important to the activities of the Association. One such person was Arthur Neville Brown (1864-1934; Figure 1), who for many years actively supported the work of the Variable Star Section (VSS) as a prolific observer and as its Secretary, engaging in the painstaking work of collating thousands of observations and preparing them for publication; he was also a long-serving member of the BAA Council. Brown was afforded four obituaries: in the BAA *Journal* (1), the *Monthly Notices of the Royal Astronomical Society* (2), the *BAA VSS Circulars* (3) and in The Observatory (4).

This paper discusses Brown's life and career, both as a variable star enthusiast and as a dedicated and highly respected schoolmaster.

**Early life and education**

A.N. Brown was born on 18 June 1864 at Nayland in Suffolk, a village 10 km north of Colchester where his father, the Reverend James Taylor Brown (1829 – 1875), was vicar of St. James' Church (5). His mother was Rachel Frobisher Jones (1837-1910) (6). In 1867 the family moved to Preston, Lancashire, where the Reverend Brown became vicar of Holy Trinity Church; he had a long association with this Church having been baptised there and later serving as a Curate. There was a strong family connexion with the north-west of England: James' father, Alderman Robert Brown (1808-1858) had been a Preston medical practitioner. One of James' brothers, Robert Charles Brown (1836-1925), was a surgeon in Preston (7), and his other brother was a vicar in Stockport, also serving as rural Dean of Macclesfield and honorary Canon of Chester Cathedral (8).

It is clear that the Reverend Brown had a challenge on his hands in his new benefice for according to a contemporary report (9) "the congregation was wretchedly thin, awfully scarce, and just on the borders of invisibility". Initially he made progress and the congregation swelled, "But he is far too good to be a parson. A gentle melancholy seems to have got hold of him". His sermons also received a mixed reception from those sitting in the pews: "He always preaches sincerely; a quiet spirit of simple unadorned piety pervades his remarks, but he depresses you too much…and his words sometimes chill like a condensation of Young's 'Night Thoughts'" (10).

The Reverend Brown died suddenly in early 1875 at the age of 45 (11). He had previously sold his house to the parish for use as a vicarage, which meant that he could carry on living there rent-free. However, upon his death his family had to vacate the house to make way for the new vicar. At that time A.N. Brown was only 10 years old, the eldest of 8 children, with one born posthumously. He was sent to the Clergy Orphan School in Canterbury (12).

According to Hugh Casement, A.N. Brown's great-nephew and Brown family historian, "the problem is that after his mother was widowed she became something of a recluse, and relationships with her children were strained. She did not provide a welcoming home for them to return to, and as a result they went their separate ways and hardly kept up with each other" (8).

After leaving school, A.N. Brown matriculated at Queen's College Oxford in the autumn term of 1883, where he won a Classical Exhibition, going on to graduate BA in 1887 and MA in 1891.

Brown remained a bachelor throughout his life.

**Schoolmaster at Ludgrove**

In 1888 Brown embarked upon his teaching career when he was appointed First Classical Master at Canon Lovett Cameron's Preparatory School at Mortimer, Berkshire. The boarding school had not long been established by the Reverend Charles John Leslie Lovett Cameron (1843-1927), vicar of St. John's Mortimer, and he remained there for 10 years.

In 1898 he moved to Ludgrove School, near Barnet in Hertfordshire (13), where he became Assistant Master. Ludgrove had been established in 1892 by the footballer Arthur Dunn (1860-1902) with the aim of preparing boys for entry to Eton and other Public Schools. Dunn had recruited a number of other eminent sportsmen to assist him as masters; Brown did not fall into this category although he had been a keen oarsman at University (14). When Dunn died prematurely in 1902 two of these masters, G.O. Smith (1872-1943) and William "Bill" Oakley (1873-1934) became joint

Headmasters. Smith and Oakley (Figure 2) had played for Dunn's old football club, the Corinthians, which was one of the best known amateur football clubs of the time, and both had captained the England team. Smith and Oakley bought the partnership from Dunn's widow, Helen. Brown also assisted in this process with financial support, thereby becoming a co-partner with Smith and Oakley, although he seems subsequently not to have played any directive or administrative role in the life of the school (15). A photograph of Brown and some of the boys from his Division, or house, is shown in Figure 3.

Given the demographic intake of Ludgrove, it is not surprising that Brown taught many boys who would later become famous. His best known pupil was Alec Douglas-Home (1903-1995), later Lord Home of the Hirsel, who was Prime Minister between 1963 and 1964. Home had fond memories of Brown: (15)

"The master to whom I and countless others were most indebted was a non-games-playing teacher of classics who for some forgotten reason we christened 'Bunco' Brown. He was rather a forbidding figure, invariably dressed in a greenish brown herring-bone tweed suit with a stick-up collar which, with a large drooping moustache and plastered-down hair, produced a portrait on unrelieved gloom. But as an interpreter of Virgil and Ovid, Homer and Plato, he was a genius. He could even bring alive the interminable campaigns of Caesar. As a politician, where above all an accurate use of language matters, I gladly acknowledge my debt to 'Bunco' Brown. He roused in me a love of learning for its own sake which has stood me in good stead ever since".

Alec Douglas-Home's younger brother, William (1912-1992), who became a playwright, also mentioned Brown in his autobiography.

Another pupil, Roland Pym (1910-2006), who went on to become a painter, illustrator and theatrical set designer, shared his vivid memories of Brown: (15)

"Bunco Brown was a master of whom I have the clearest memory – he was very strict and quite a frightening-looking person with a huge walrus moustache and a very old-fashioned 1870s-style suit of thick ginger brown tweed with a high stiff collar; he looked a bit like Crippen the murderer, and rumour had it that he was once arrested because of it! He pronounced 'Breakfast' like 'Breaf-kast' which we always tried to make him say so that we could laugh about it. He was a brilliant teacher of classics, and was also a very keen astronomer with a large telescope which he kept in the Master's Cottage nearby, overlooking the school yard. One day in the yard I did an imitation for my friend William Douglas-Home of Bunco Brown looking through his telescope – and then suddenly realised to my horror that he was looking right down at me, he'd been watching the whole thing from the garden next door. I got hauled in by him and punished for it in some way, I don't remember how".

In addition to teaching Classics, Brown ran extra-curricular courses in astronomy for the boys which were well attended, but lest they be thought of purely as fun, they concluded with an examination. He also gave lectures on other subjects details of which are recorded in the *School Notes*. He often spoke about his overseas travels since he travelled widely in the school holidays, during which his knowledge of French and German sometimes came in handy. Destinations included Switzerland, French North Africa (16), Greece, the United States and Canada, including a journey across the country on the Canadian Pacific Railway. He also spoke about his trip to the Balearic Island of Majorca to view the total solar eclipse on 30 August 1905. There he observed and photographed the eclipse from just outside Palma, along with many BAA members. He recalled visiting the eclipse camp of Sir Norman Lockyer and seeing HMS *Venus* anchored in Palma harbour which was present to lend general support to the eclipse observers as well as providing them with time signals (17).

In 1900 Brown established a branch of the *Navy League* at Ludgrove. The aim of the *League* was to promote awareness amongst the British public, especially young people, of the dependency of the country on the sea and that the only safeguard was to have a powerful navy. Sometimes Naval officers came to the school, but other times Brown spoke himself and he was said to have a good knowledge of strategic naval matters, often describing the details of historic maritime battles such as Trafalgar. His 1904 talk on "Our Fleet Today" was prophetic; he drew attention to the rise of Germany as a naval power, which at the time was becoming a national concern. He concluded with the hope that "it will be a very long time before our ships have to confront an enemy in battle", and that the Royal Navy might continue for many years its peaceful patrol of the oceans that surround the shores of the British Empire" (15). Sadly, within 14 years no fewer than 15 of the boys listening to the lecture would lose their lives in the First World War. One such was John Spencer Dunville, VC (1896-1917). In June 1917, while Dunville was serving in the First Royal Dragoons, he died from wounds he received at Epehy in France. He had been protecting an N.C.O. of the Royal Engineers who was cutting barbed wire which had been laid by the enemy. Although he was wounded by the enemy's fire, he continued to direct his men until the wire-cutting operation had been successfully completed. He remained conscious, but died from his wounds the next day. The Victoria Cross was awarded posthumously (18).

**The BAA and variable star observations**

Brown's interest in astronomy began in about 1891, but it didn't really take off until 1905 when he purchased a 5-inch (12.5 cm) Watson-Conrady refractor on an altazimuth mount. This was soon supplemented with a 3-inch (7.5 cm) refractor (19) and both instruments can be seen in Figure 1. The first object he observed with the Watson-Conrady was Saturn on 30 September 1905. This was followed by clusters, double stars and nebulae. Throughout his observing career he never had an observatory and used the telescopes in the open air, moving them around the sites from which he observed to obtain the best view.

With a growing interest in astronomy, Brown was elected to the BAA on 29 November 1905, becoming a life member (20).  He was elected Fellow of the RAS on 12 April 1907. Almost immediately upon joining the BAA he contacted the Director of the VSS, E.E. Markwick (1853-1925) expressing a wish to undertake a systematic programme of observing variable stars. Markwick had been Director since 1899 and was a popular and enthusiastic leader of the Section (21) and he soon took Brown under his wing. Brown's first variable star estimate was made on the evening of 23 February 1906 (1) and was of the well-known (and easily located) Long Period Variable (LPV) R Leo. This was to be the first of 557 observations of the star, which continued until May 1934, and the first of a total of about 40,000 variable star estimates (22). Almost immediately Brown established himself as one of the most prolific observers in the VSS. Markwick noted in his VSS Director's report for the 1905/1906 session that "An excellent series of observations of long-period variables has already been made by Mr. Brown" (23). He expanded enthusiastically upon this in his annual report on LPV observations for 1906: (24)

"There are been an immense infusion of energy into the Section during the year 1906, the total number of observations, 2,865, being more than three times that in 1905. Out of these Mr. A.N. Brown contributes 950, just about one-third; and the quality of his work is as good as the quantity. Both he and P.M. Ryves [with 580 observations] have made some very fine series of light-determinations; in several cases, the light curves can well be deduced from the work of these two energetic observers taken separately. They have not hesitated to prosecute their researches far into the early morning hours, whereby observations of a star are obtained which are well distributed over the year".

Whilst the increase in the number of VSS observations was clearly welcome, it also meant that additional clerical effort would be entailed in the painstaking activity of collating, analysing and reporting of the Section's work. Markwick noted: (24)

"if progress continues at this rate, it will be necessary to distribute some of the work of dealing with such a large mass of observations among a few of the Members. Fortunately, help has been freely given in the past, and will no doubt be forthcoming in the future".

Almost immediately Brown responded by volunteering to help with compiling observations and he made a major contribution to the preparation of the Section's *Memoir* on LPVs covering the years 1905-1909, which was eventually published in 1912 (25), and the following four *Memoirs* (26). This meticulous work involved generating lists of the observations gleaned from individual member's reports, checking the estimates and derived magnitudes, compiling the observations in chronological order. Only then could they be put in final manuscript form. Even after submitting the manuscript he had to read, cross-check and revise the printer's proofs. Whilst he did not do this single-handedly, he did the bulk of the work. In this manner a total of some 160,000 observations went across his desk. As future Section

Director, Félix de Roy (1883-1942), remarked (1) "[t]hose variable star observers whose work passed through his hands owe him a large debt for the full justice which was done in this way to their observations".

As well as his involvement with the production of the VSS *Memoirs*, Brown also made generous financial contributions to their publication. For example, in 1910 he donated £50 towards the LPV *Memoir* covering 1905-1909. In 1908 he had contributed £20 to defray the costs of producing blueprints of VSS light curves so that they could be shared with astronomers around the world including Harvard College Observatory (HCO), which was one of the global centres for variable star research.

When Markwick retired as VSS Director at the end of 1909, he was succeeded by Charles Lewis Brook (1855 – 1939) (27). Brook relied on Brown for compiling observations and clerical work just as much as Markwick had done. When Brook relinquished the Director's position at the end of 1921 it was natural that Brown was offered the Directorship, given his extensive practical knowledge of variable stars and his involvement with the Section's activities over many years. However, Brown always preferred working in the background and he was by nature a private person, thus after due consideration he declined the offer. The position was offered to Félix de Roy, a Belgian national living in Antwerp. De Roy was also well-known to Section members as an active observer, having joined the Section at about the same time as Brown. He had also taken refuge in London during the First World War and had attended many BAA meetings so was also known widely in the Association (28). Since de Roy was living in Belgium, he appointed Brown as his representative in Britain, as well as Secretary. In this capacity Brown was to 'receive at his address all observations of members of the Section residing in the British Empire' (29), to distribute charts and observation forms, to assist new observers and to archive original observations. In practise, Brown essentially ran the section on a day-to-day basis, receiving observations and compiling them, corresponding with members, especially new ones to whom he gave advice. De Roy was always generous in recognising publicly Brown's contribution: (29)

"Most of the [VSS] Members will…have been…impressed by his kindness and helpfulness in dealing with all matters submitted to him in the course of his exacting duties; by his prompt, always courteous and encouraging replies to the letters and inquiries which were sent to him from all sides, and by the lovable character of the man which his correspondence betrayed".

**Life in retirement**

Brown retired from Ludgrove School in 1923. He moved to a house, *Brackenhurst*, at Bucklebury Common, Berkshire, about 18 km west of Reading, not too far from where he lived whilst teaching at Canon Lovett Cameron's Preparatory School. He became a keen gardener and welcomed many visitors at *Brackenhurst*, including de Roy during his trips to England. His other interests included bird watching and music; he was an accomplished pianist and organist and was intensely interested in the work of Cecil Sharp (1859-1924), the founding father of the revival of English folk songs in the early twentieth century. He also corresponded with the musicologist, A.H. Fox Strangways (1859-1948).

The timing of Brown's retirement was opportune as it allowed him to devote even more time to the VSS Secretary duties that he had recently taken on. De Roy gave him the title of *Honorary Secretary* in 1924 as a mark of appreciation. As the years went on, Brown took on more of the Section's work as de Roy's work commitments as Editor of a Belgian newspaper became more onerous.

Brown also gained international exposure. He became a member of the IAU Commission 27 on Variable Stars in 1925. Then in 1928, as a result of his acknowledged contribution to variable star astronomy, Brown was made an Honorary Member of the American Association of Variable Star Observers (AAVSO) (30). Since becoming VSS Secretary he had regularly exchanged letters with the AAVSO's Recorder, Leon Campbell (1881-1951). He also arranged for copies of the VSS Memoirs to be sent to both Campbell and HCO. Over the years they discussed comparison star sequences for certain variable stars, as well as ways of increasing the cooperation between the two variable star organisations (31). One idea that had been floated by de Roy is that the two organisations should divide the more popular stars between them so as not to duplicate. However, whilst he was willing to entertain the idea, Campbell's main objection (32), which he shared with Brown, was the delay in publishing the BAA's results, which meant that it took time before they were available to other researchers. The *Memoirs* covered 5-year intervals and were published several years in arrears owing to the time taken to compile them. By contrast AAVSO data were published monthly. On the other hand, Campbell's objection might not have been completely justified as many interim reports were published in the BAA *Journal*.

Campbell also engaged Brown to observe some specific AAVSO stars that were not on the BAA's programme (33). In a similar vein, Brown, acting on a request from W.H. Steavenson (1894-1975), asked for a list of faint, or otherwise less well observed stars, that Steavenson could profitably observe with his 18-inch (46 cm) reflector (34).

As well as his variable star work, Brown served on the BAA Council many times, between 1916 until his death in 1934. He was also the Association's representative on the National Committee for Astronomy from 1931. He regularly travelled to London to attend Council meetings. However, he did not often attend other BAA meetings, apart from the infrequent VSS Section meetings (35), partly because of his modest and retiring personality and partly because he viewed meetings as interruptions to his observing and compiling routine (2).

Although Brown was intensely interested in the observational aspects of variable stars, and his introductory guide which he co-authored with de Roy would be useful even to today's visual observer (36), he was less interested in the astrophysical consequences.

**Mary Adela Blagg and the IAU**

In 1928 Brown travelled to Leiden in the Netherlands to take part in the Third General Assembly of the IAU which was held between 5 and 13 July. He attended meetings of Commission 27 on Variable Stars along with Mary Adela Blagg (1858-1944; Figure 4), who was also a member of the Commission (37). Blagg, an amateur astronomer who was known for her work on lunar nomenclature, had become interested in variable stars though Prof H.H. Turner (1861-1930) of Oxford University who had asked for volunteers to analyse the variable star observations of Joseph Baxendell (1815-1887). Together Blagg and Turner authored ten papers on the results, which appeared in *Monthly Notices of the Royal Astronomical Society,* although Turner credited her with having done most of the work. Blagg was an unassuming woman who never married; she seldom left her native Cheadle in Staffordshire and was rarely seen at meetings, exceptions being the IAU meetings at Cambridge, UK, in 1925 and Leiden (38). She also analysed observations of other variable stars bringing to bear her considerable mathematical skills, which were largely self-taught. Brown drew to her attention his long term observations of 3 LPVs, which were outside the VSS programme, and she set about analysing them and published the results in three papers in *Monthly Notices.*

The first star Blagg investigated was RT Cyg, in which she considered Brown's observations made between 1908 and 1928 (39). Brown's own light curves from the earlier part of this period are shown in Figure 5. Blagg carried out a thorough mathematical analysis of 40 cycles of the star and found a mean period of 191.2 days, with a near sinusoidal variation, with the rise being slightly more rapid that the decline. She determined the equation for the shape of the light curve and determined the dates of maximum and minimum. An analysis of the O-C residuals suggested a possible half-yearly variation. A similar analysis of V Cas between 1909 and 1929 followed (40), revealing a mean period of 228.76 days. The final LPV examined was U Per, which she found had a mean period of 320 days in the interval 1909 to 1930 (41).

*Sic itur ad astra* (42)

Brown's variable star work continued unabated until he fell ill in the middle of October 1934. He made his last variable star observations on 12 October (43) and he was still at his desk on 19 October dealing with VSS matters and BAA Council business, for he was still a member of Council. He passed away peacefully at home at Brackenhurst early on 4 November 1934 with heart failure. His funeral was held on 7 November at the Church of St. Mary the Virgin, Bucklebury, which was attended by several BAA Council members and the Astronomer Royal and BAA President, Harold Spencer-Jones (1890-1960). His body was interred in the churchyard (Figure 6). At the following BAA meeting tributes were paid to him by W.H. Steavenson, F.M. Holborn and Spencer-Jones, who commented that "Mr. Brown was a man who did a large amount of excellent work in a very unobtrusive manner and the Association had been much indebted to him" (44). The following year Brown's Executors presented some of his books to the BAA Library.

After so many years of dedicated service Brown was sorely missed by the VSS. His position as Secretary, and as BAA Council member, was assumed by W.M. Lindley (1891-1972), who eventually went on to become Director (45). As de Roy, who had known Brown well for many years, noted in Brown's obituary: (1)

"A.N. Brown was a charming personality, unassuming almost to excess, and endowed with a beautiful and noble sense of duty. To all who knew him, he was the best and most faithful of friends. He will be remembered as a very distinguished amateur astronomer".

It is upon the foundations laid by dedicated people such as A.N. Brown that the BAA continues to build even today.

**Acknowledgements**

Many people have been very helpful to me in this research. Wendy Spackman of Ludgrove School, answered many questions and spent considerable time delving into the Ludgrove School archives for information about Brown. She kindly provided copies of papers and the photograph of Brown's Division used in this paper, the first photograph of Brown to appear in a BAA publication. Hugh Casement, Brown's great-nephew, provided many helpful details about Brown's background and family history from his own research. John Toone brought the photograph of Brown used in Figure 1 to my attention. Julia & Keld Smedegaard allowed me to use their photographs of Brown's grave (and put me in touch with Hugh Casement). Mike Saladyga searched the AAVSO archives and provided copies of the correspondence between Brown and Leon Campbell, AAVSO Recorder. Geoffrey Culshaw tracked down the photograph of Mary Adela Blagg, which is the only photograph of her of which I am aware. Richard Baum was the original inspiration for initiating this research and I thank him for the valuable and enjoyable discussions I have had with him on many historical aspects of astronomy.

This research made use of scanned back numbers of the JBAA, a wonderful resource the existence of which is largely thanks to the efforts of Sheridan Williams, SIMBAD, operated through the Centre de Donées Astronomiques (Strasbourg, France), and the NASA/Smithsonian Astrophysics Data System.

Finally I thank the referees Dr. Mike Leggett, Roger Pickard and Mike Frost for their helpful comments that have improved the paper.

**Address**

"Pemberton", School Lane, Bunbury, Tarporley, Cheshire, CW6 9NR, UK [bunburyobservatory@hotmail.com]

**Notes and references**

1. de Roy F., JBAA, 45, 323-326 (1935).

2. Lindley W.M., MNRAS, 95, 318-319 (1935).

3. de Roy F., BAA VSS Circular 10, 20 November 1934.

4. Observatory, 57, 387 (1934).

5. The Reverend James Taylor Brown was vicar of Nayland from 1863 to 1867. Nayland is in Dedham Vale, in "Constable Country", and the church contains the famous painting, *Christ Blessing the Bread and Wine*, by John Constable (1776-1837). James was born in Preston on 23 December 1829 and baptized at Holy Trinity, Preston, on 10 February 1830.

6. Rachel's father was the Reverend Neville Jones, vicar of St. George, Little Bolton (Bolton le Moors).This is where A.N. Brown's middle name came from. Rachel and James married on 27 August 1863 at St. George's, where James was curate.

7. Sir Robert Charles Brown, MB, MA, FRCS, FRCP was instrumental in raising funds for building the Preston Royal Infirmary; he was also a benefactor in his own right; biographical entry in the Royal College of Surgeon's *'Plarr's Lives of the Fellows online*'. http://livesonline.rcseng.ac.uk/biogs/E000997b.htm.

8. Casement H., Personal communication (2012). Other siblings were Harold Holgate Brown (newspaper correspondent during the Boer War), Robert Charles Brown (medical doctor), Richard Cecil Brown (Indian Civil Service), James Clement Brown (London solicitor), Emily Edith Mary Brown, Francis Julia Brown, Henry Martyn Brown (the posthumous son, Colonel in the Indian Medical Service).

9. The following quotes are taken from The Church of the Holy Trinity in the Town of Preston in "*Our Churches and Chapels*" by A. Hewitson which was printed at the "Chronicle" Office, Fishergate, Preston (1869).

10. The poem "*Night Thoughts on Life, Death and Immortality*" was published in 1742 by Edward Young (1681-1765).

11. James died at 27 Winckley Square, Preston. That was not his home, which was on the other side of the Square, but where his mother and his brother Charles lived. The obituary in the Preston

Guardian said he died of "brain disease". Whether this was a brain tumour, embolism, aneurysm or something else is not known.

12. The school is now St. Edmund's School, Canterbury. Two other brothers attended the same school. Another two attended Christ's Hospital school in London. Casement H., Personal communication (2012).

13. In 1937 the school was moved to its present location at Wixenford, Wokingham.

14. *Cricket Archive Oracles* ( http://www.cricketarchive.com) lists Brown as having played for Lincolnshire against Staffordshire at Stoke-on Trent on 1 and 2 August 1890. Although it identifies "our" A.N. Brown through his birth date and location, I have no other evidence that he played cricket at the county level.

15. Barber R., *The Story of Ludgrove*, Guidon Publishing (2004).

16. Algeria, former French Morocco and Tunisia.

17. Levander F.W. (ed.), Mem. Brit. Astron. Assoc., The Total Solar Eclipse of 1905. BAA (1906).

18. Holroyd J. & Holroyd M., The Dumvilles of Hunton and Other Dumville, Dunville and Domville Families (2012) http://www.dumville.org/photo_pages/jd1896_pho.html.

19. The present whereabouts of Brown's telescopes is not known. The author would be interested in hearing from anyone who has knowledge of them.

20. Brown's proposer for BAA membership was Henry T.C. Knox (d. 1924) and seconder was A.C.D. Crommelin (1865-1939). Crommelin was Director of the BAA Comet Section 1897-1939. Candidates for election as members of the Association on 29 November 1905: JBAA, 16, 48 (1905). New members of the Association elected on 29 November 1905: JBAA, 16, 82 (1905).

21. A biography of Markwick by the present author is available: Shears J., JBAA, accepted for publication (2011), http://arxiv.org/abs/1109.4234 .

22. The number of 40,000 observations is mentioned in the obituaries by de Roy and Lindley. The BAA VSS online database lists a total of 30,293 variable stars estimates by Brown (Observer Code BN). This is an underestimate as it is thought that not all of his observations have been added yet. Moreover, his observations of some stars for the AAVSO were not included in the BAA VSS database.

23. Markwick E.E., JBAA, 16, 384 (1906).

24. Markwick E.E., JBAA, 17, 343 (1907).

25. Markwick E.E., Mem. Brit. Astron. Assoc., 18 (1912).

26. Mem. Brit. Astron. Assoc., 22 (1918), 25 (1925), 28 (1929) and 31 (1934).

27. C.L. Brook's biography can be read at: Shears J., JBAA, 121, 17-30 (2012).

28. A biography of de Roy can be read at: Shears J., JBAA, 121, 203-214 (2011).

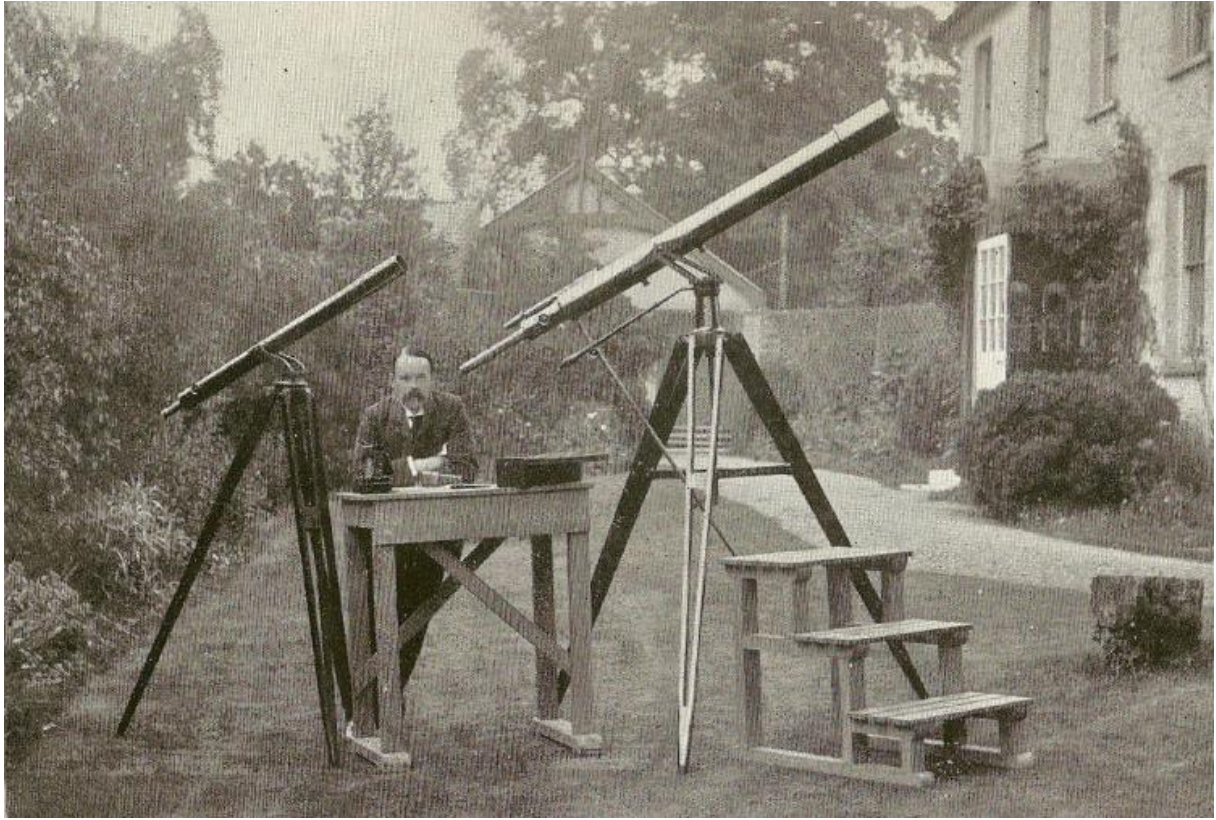

Figure 1: A.N. Brown (1864-1934) in the garden at Ludgrove School with his 5-inch and 3-inch refractors. 28 July 1909. From reference (1).

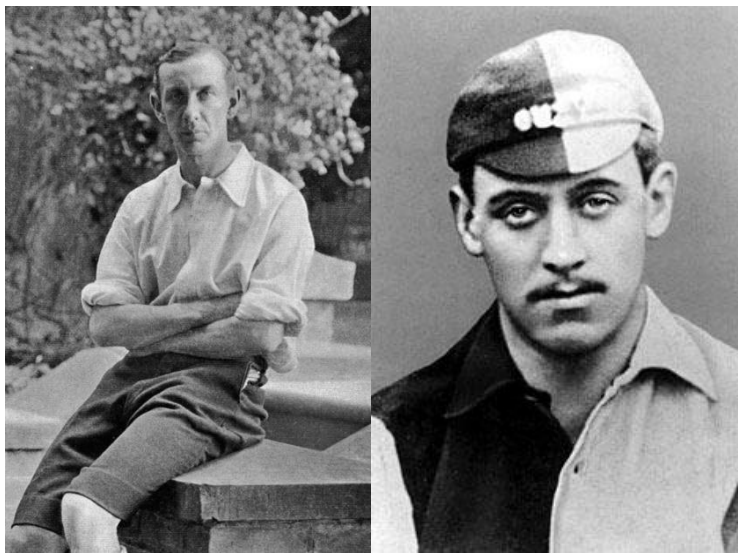

Figure 2: *Left* G.O. Smith (1872-1943) and *right* William Oakley (1873-1934), masters at Ludgrove School

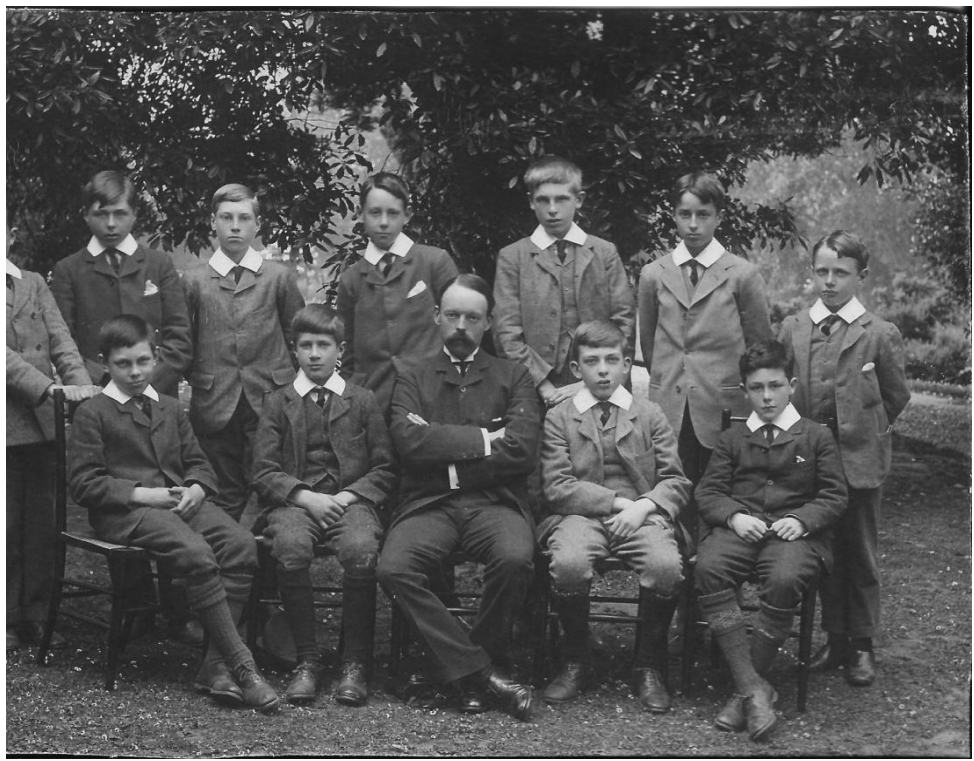

Figure 3: A.N. Brown and boys from his Division

(Ludgrove School)

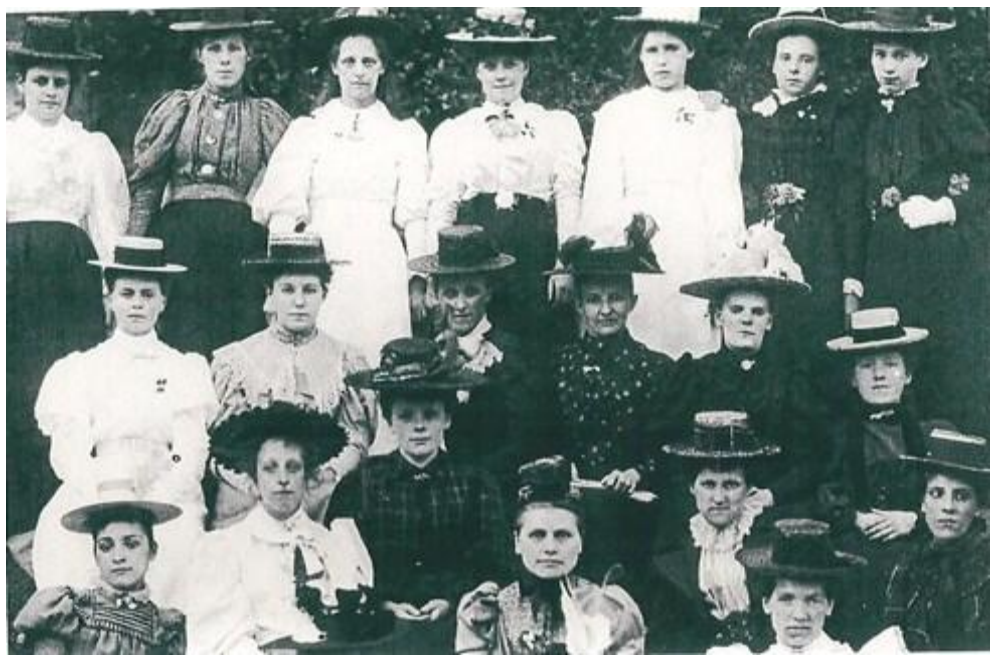

Figure 4: Mary Adela Blagg (1858-1944) and members of the Cheadle Girls Friendly Society, ca.1890

Blagg is at the bottom in the centre. (Geoffrey Culshaw)

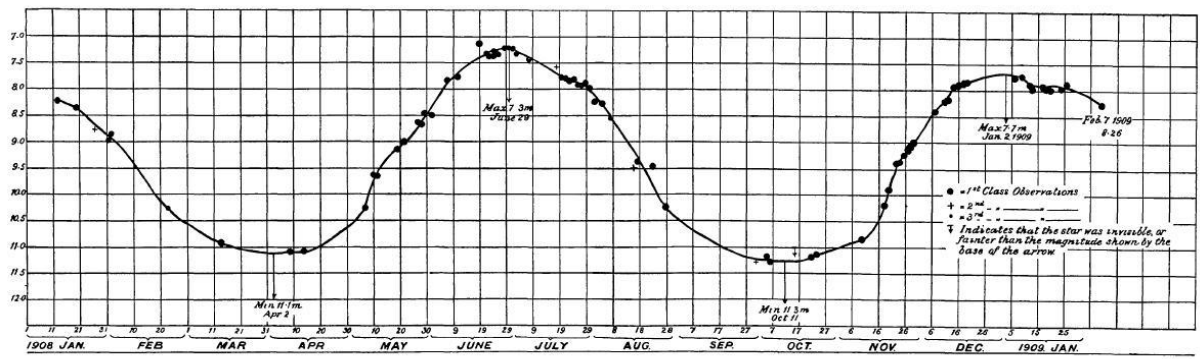

Light Curve of RT Cygni in 1908.—A. N. Brown.

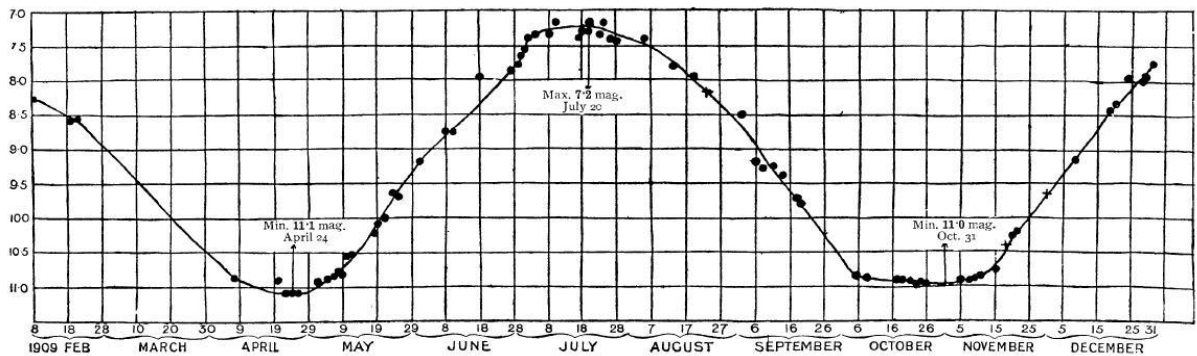

Light Curve of RT Cygni in 1909.—A. N. Brown.

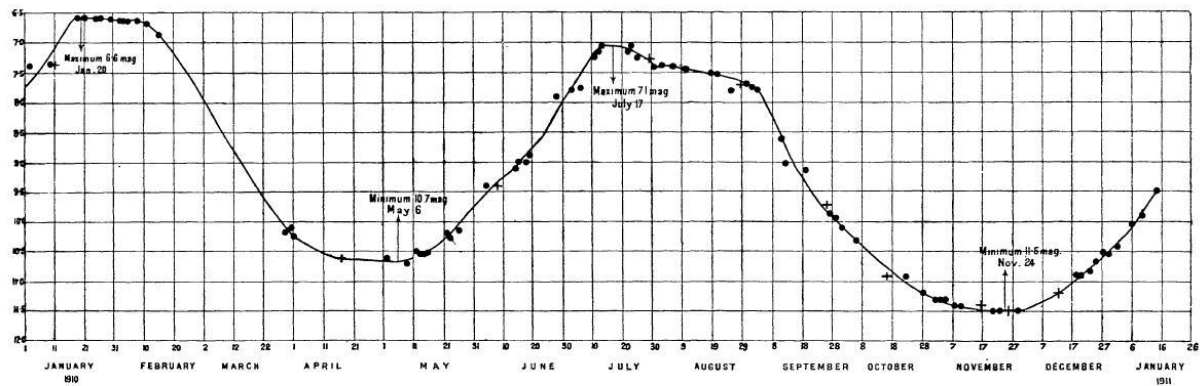

Light Curve of RT Cygni, 1910.—A. N. Brown.

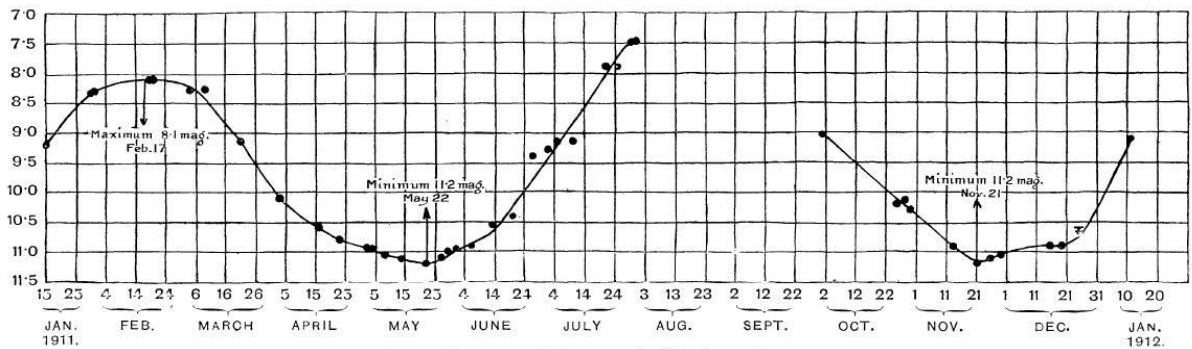

Light Curve of RT Cygni, 1911.—A. N. Brown.

Figure 5: Brown's observations of RT Cyg 1908-1911 (46)

Note that the scales are slightly different in each plot

*Accepted for publication in the Journal of the British Astronomical Association*

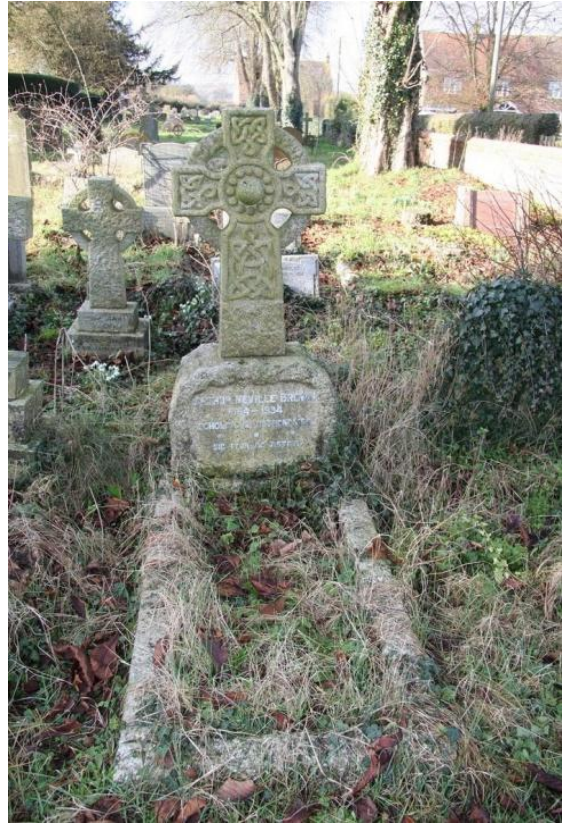

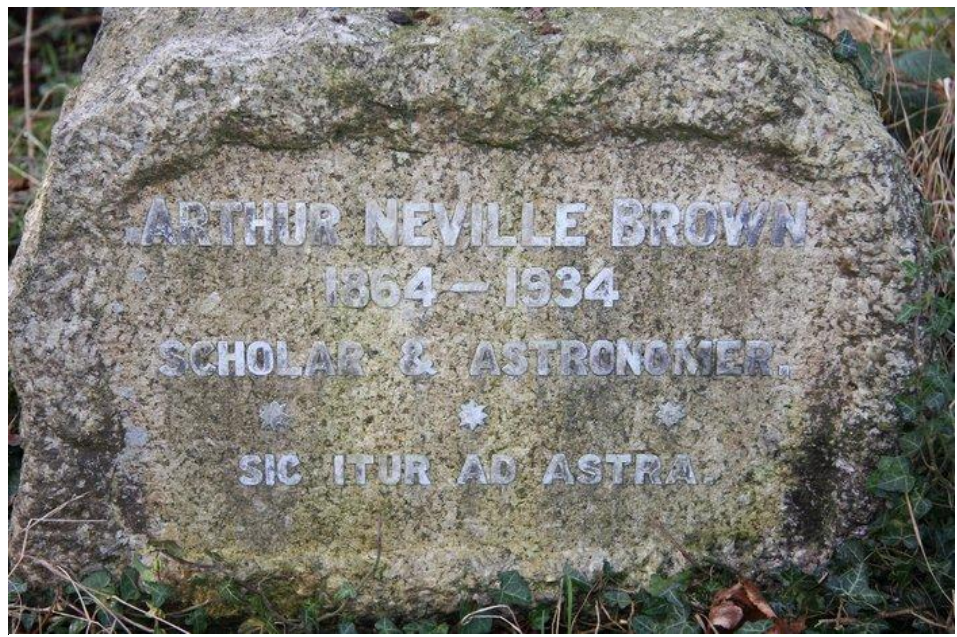

Figure 6: Brown's grave and headstone at Bucklebury

(Julia & Keld Smedegaard, February 2011)

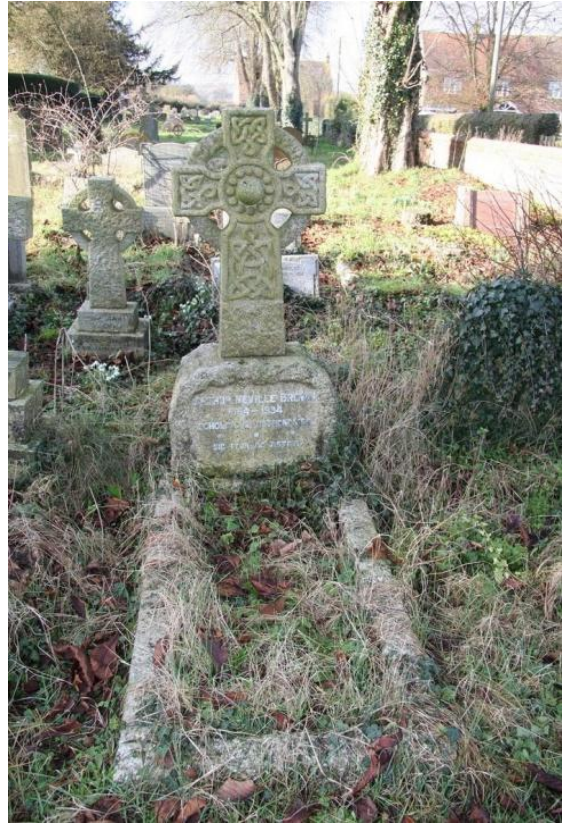

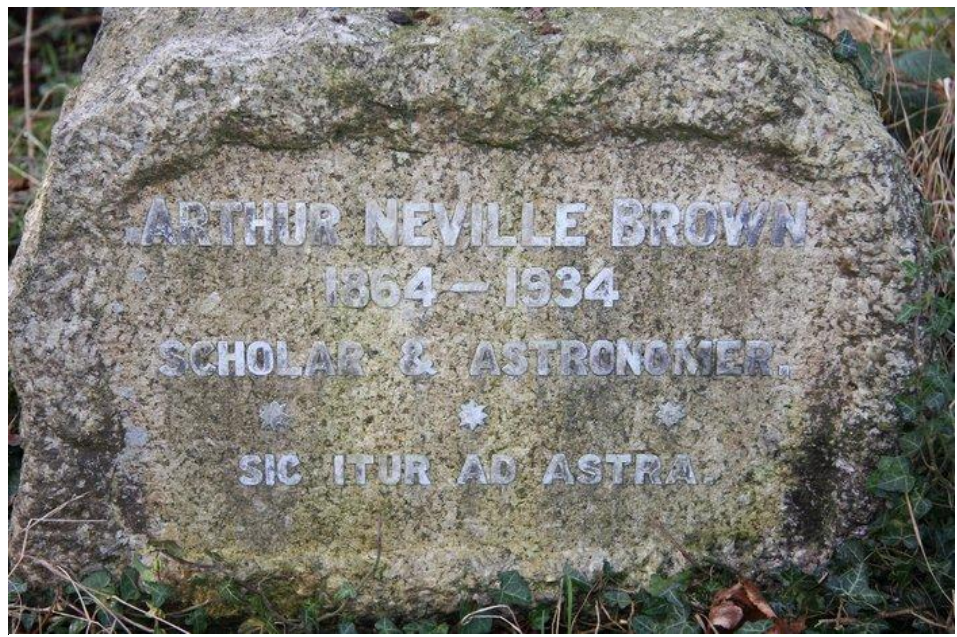

Figure 6: Brown's grave and headstone at Bucklebury

(Julia & Keld Smedegaard, February 2011)

*Accepted for publication in the Journal of the British Astronomical Association*